\documentstyle[prc,aps,epsfig,multicol]{revtex}
\tightenlines
\newcommand{\bm}{\bibitem}

\def\be {\begin{equation}}
\def\ee {\end{equation}}
\def\bea {\begin{eqnarray}}
\def\eea {\end{eqnarray}}

\def\ks {{|{\vec k}|}^\ast}

\def\mn {m^{*}_n}
\def\kb {\bf k}
\def\2l {\frac{{f_i}}{(2\lambda + 1)}}
\def\ks {k\!\!\!/}

\def\var {\bf}

\begin{document}
\title{Matter induced $\rho-\delta$ mixing : a source of dileptons }
\author{ O. Teodorescu, A. K. Dutt-Mazumder, and C. Gale  }
\address{ Physics Department, McGill University\\ 3600 University St., 
Montreal, Quebec H3A 2T8, Canada\\}
\maketitle

\begin{abstract}
We study the possibility of $\rho-\delta$ mixing via N-N excitations in 
dense nuclear matter. This mixing induces a peak in the
dilepton spectra at an invariant mass equal to that of the $\delta$.
We calculate the cross section for dilepton production through the
mixing process and we compare its size with that of
$\pi-\pi$ annihilation. In-medium 
masses and mixing angles are also calculated.
\end{abstract}
\vspace{0.3 cm}

%Keywords: quantum hadrodynamics,
%vector-scalar mixing, dilepton spectra
%
\vspace{.2 cm}
%\begin{multicols}{2}

Heavy ion physics has recently seen a considerable effort 
being devoted to the study of 
the properties of hadrons in a hot and/or dense 
nuclear medium. Those activities were stimulated in part by the 
suggestion 
that in the nuclear medium, the vector meson masses would drop from their
values in free space and that this could be interpreted 
as a precursor phenomenon of chiral symmetry restoration\cite{brown91}. 
Several attempts have been made  to highlight and understand the
in-medium behaviour of vector mesons, both in theory and 
experiment \cite{brown91,ceres,lkb96,rcw97}.  
In this respect electromagnetic signals constitute valuable probes,
especially lepton pairs. 
This owes to the fact that the leptons couple 
to hadrons via vector mesons and therefore hadronic processes involving 
$e^+e^-$ in the final channel are expected to reveal their properties in
the dilepton spectra. 
Furthermore, the $e^+e^-$ pairs suffer minimum final state
interactions and are thus likely to bring information to the detectors 
essentially unscathed. 
Several experiments have measured, or are planning to measure,  the lepton 
pairs produced in
nucleus-nucleus collisions. They have been carried out by  
the DLS at LBL \cite{dls}, and by 
HELIOS \cite{hel} and 
CERES \cite{ceres} at CERN. Two new initiatives that will focus on
electromagnetic probes will be PHENIX at RHIC \cite{phenix} and HADES at 
GSI \cite{hades}.  
The density-dependent characteristics of vector mesons can also be
highlighted through experiments performed at TJNAF \cite{tjnaf}.
The last two projects will involve measurements performed
in environments where the possible modifications from vacuum properties
will mostly be density-driven. It is with those in mind that we have
performed the theoretical estimates about which we report in this paper.  

%are primarily geared to bring out density dependent characteristics of
%the vector mesons. In fact in the latter case the bombarding energy being
%low compared to the former ones, the temperature is not supposed to 
%play any significant role. In the present calculation we confine ourselves
%to the case relevant for low energies.
%In this context it might be mentioned that there have already been claims that 
%in dense matter vector meson masses indeed undergo a 
%reduction compared to their values in vacuum \cite{jin95,altemux}. 

While several theoretical studies have sought to investigate the
in-medium properties of the vector mesons (mainly the $\rho$), their 
possible mixing with other mesons has only started to receive attention
in the context of dense baryonic matter. An exception is the 
case of $\rho$-$\omega$
\cite{abhee97,broni98}. This specific mixing can be omitted when 
dealing with symmetric 
nuclear matter, as we will here.
%\footnote{ A marginal $\rho$-$\omega$ or $\rho$-$\sigma$ 
%mixing can of course occur arising out
%of the neutron proton mass difference \cite{abhee97} }
The popularity of the $\rho$ meson resides
in the fact that in nuclear collisions a substantial contribution to the
dilepton spectra comes from
$\pi$-$\pi$ annihilation which proceeds through $\rho$ as an intermediate
state.  This fact can also be stated as the dilepton spectrum sampling
the in-medium vector meson spectral function \cite{gaka91}.

We explore here the possibility of $\rho$-$\delta$ 
(or $a_0$ as listed in \cite{pdb}) mixing via nucleon(n)-nucleon(n)
excitations in nuclear matter.  Such a mixing, in effect, is similar 
to the known $\omega$-$\sigma$ mixing \cite{chin77,walecka86,saito98,wolf98}. 
This is a pure density-dependent effect and is forbidden in free space on 
account of Lorentz symmetry. We will show that such a mixing opens
up a new channel for the dilepton productions and induces an additional peak 
in the $\phi$ mass region.  

The interaction Lagrangian we will use can be written as 
\bea
{\cal L}_{int} = g_\sigma {\bar \psi}\phi_\sigma \psi + 
          g_\delta {\bar \psi}\phi_{\delta,a}\tau^a \psi
          + g_{\omega NN}{\bar{\psi}} \gamma _\mu\psi\omega^\mu
          + 
 g_{\rho} [{\bar{\psi}} \gamma _\mu \tau^\alpha
 \psi + \frac{\kappa _\rho}{2m_n}{\bar{\psi}}
    \sigma_{\mu\nu}\tau^\alpha \partial ^\nu] \rho^\mu_\alpha\ ,
\eea                                                 
where $\psi$, $\phi_\sigma$, $\phi_\delta$, $\rho$ and $\omega$ correspond
to nucleon, $\sigma$, $\delta$ , $\rho$ and $\omega$ fields, and $\tau_a$ 
is a  Pauli matrix. The values used for the coupling parameters are obtained 
from Ref. \cite{mach89}.

The polarization vector through which the $\delta$ couples to $\rho$
via the n-n loop is given by 
\bea
\Pi_ \mu (q_0,|{\vec q}|) &=& 2i g_\delta g_\rho \int \frac{d^4k}{(2\pi)^4}
    \mbox{Tr}[G(k) \Gamma_\mu G(k+q)]. \label{pim}
\eea
where 2 is an isospin factor and the vertex for $\rho$-nn
coupling is:
\bea
\Gamma_\mu=\gamma_\mu - \frac{\kappa_\rho}{2m_n}\sigma_{\mu\nu}q^\nu\ .
\label{vertex}
\eea
$G(k)$ is the in-medium nucleon propagator
given by \cite{sewal}
\bea G(k_0,|{\vec k}|) = G_F(k) + G_D(k_0,{\vec k})
\label{nuclprop}
\eea
with
\bea
G_F(k)=\frac{(\ks+\mn)}{k^2-m_n^{* 2} + i\epsilon}\ ,
\eea
and
\bea
G_D(k_0,|k|)=(\ks+\mn)\frac{i\pi}{E^*_k} 
\delta(k_0-E^*_k) \theta(k_F - |\kb|)\ ,
\eea
where $E^*_k=\sqrt{k^2+m_n^{* 2}}$. The second term ($G_D$) deletes on-mass
shell propagation of nucleons having momenta below the Fermi momentum $k_F$.
In Eq.~(\ref{nuclprop}), the subscripts $F$ and $D$ refer to the free and 
density-dependent part of the propagator. In the subsequent equations
$m_n^*$ denotes the effective nucleon mass evaluated at the mean field
level \cite{sewal}. 

With the evaluation of the trace and after a little algebra Eq.~(\ref{pim}) could
be cast into a suggestive form: 
\bea
\Pi_\mu(q_0,|q|)=\frac{g_\rho g_\delta}{\pi^3} 2q^2 (2m_n^*-
\frac{\kappa q^2}{2 m_n})
\int_0^{k_F}\frac{d^3k}{E^*(k)}
\frac{k_\mu - \frac{q_\mu}{q^2}(k\cdot q)}{q^4 - 4 (k\cdot q)^2}\ .
\label{mixamp}
\eea
This immediately leads to two conclusions. First, 
it respects the current conservation condition, {\em viz.} 
$q^\mu\Pi_\mu=0=\Pi_\nu q^\nu$. Secondly, there are only two components which
would survive after the integration over azimuthal angle. In fact this
guarantees that it is only the longitudinal component of the $\rho$ meson
which couples to the scalar meson while the transverse
mode remains unaltered. Furthermore, current conservation implies that
out of the two non-zero components of $\Pi_\mu$, only one is independent. 
It should be noted here that the tensor interaction, as evident from 
Eq.~(\ref{mixamp}), inhibits the mixing.

In presence of mixing the combined meson propagator might be written
in a matrix form where the dressed propagator would no longer be diagonal: 
\bea
{\cal D} = {\cal D}^0 + {\cal D}^0\Pi{\cal D}\ .
\label{dressprop}
\eea
It is to be noted that the free propagator is diagonal and has the form 
\be
{\cal D}^0 = \pmatrix{
    D^0_{\mu \nu} & 0\cr
    0 & \Delta_0
}\ .
\label{free}
\ee
In Eq.~(\ref{free}) the noninteracting propagator for the $\delta$
and $\rho$ are given respectively by
\bea
\Delta_0(q) &=& \frac{1}{q^2 - m_\delta^2 + i\epsilon},
\label{frees} \\
D^0_{\mu \nu}(q) &=& \frac{-g_{\mu \nu}+ {\frac{q_\mu q_\nu}{q^2}}}{
  q^2 - m_\rho^2 + i\epsilon}, \label{freerho}
\eea
The mixing is characterised by the polarization matrix which contains
non-diaginal elements 
\be
\Pi = \pmatrix{
    \Pi^\rho_{\mu \nu}(q) & \Pi_\nu(q)\cr
    \Pi_\mu(q) & \Pi^\delta(q)}\ .
\label{pol}
\ee
In the above expression, $\Pi^\delta$ and $\Pi^\rho_{\mu\nu}$ refer to
the diagonal self-energies of the $\delta$ and $\rho$ meson induced by 
the n-n polarization:
\bea
\Pi^\delta(q_0,|{\vec q}|) 
&=& -2ig_\delta^2 \int \frac{d^4k}{(2\pi)^4}\mbox{Tr}[G(k)G(k+q)]
\label{pidel} \\
\Pi^\rho_{\mu \nu}(q_0,|{\vec q}|) &=& -2ig_\rho^2 \int \frac{d^4k}{(2\pi)^4}
    \mbox{Tr}[G(k) \Gamma_\mu G(k+q) {\tilde\Gamma}_\nu] \ .
\label{pirho} 
\eea

It should be mentioned that, unlike mixing, 
Eqs.~(\ref{pidel}) and (\ref{pirho}) also involve a 
free part stemming from the $G_F(k)G_F(k)$ combination which is divergent. This
therefore needs to be regularized. The regularization condition we 
employ is $\partial^n\Pi^F(q^2)/\partial(q^2) ^n\vert _{m^\ast_n\rightarrow m,
q^2=m_v^2}=0~ (n=0,1,2...,\infty )$ \cite{abhee97}. 
For $\rho$ the results may be found in Ref.\cite{abhee97} which we do not 
present here and for $\delta$ the free part of the self-energy is
given by :
\bea
\Pi^\delta(q^2)=\frac{3 g_\delta^2}{2\pi^2}[3(m^{* 2}_n-m^2_n) 
-4 (m^*_n - m_n)m_n
-(m^{* 2}_n-m^2_n)\int_0^1 dx\hspace{0.1cm} \ln \left[ \frac{m^{* 2}_n - 
x(1-x)q^2}{m^2_n} \right] \nonumber\\
-\int_0^1 dx\hspace{0.1cm} (m^2_n-x(1-x)q^2)\ln \left[ \frac{m^{* 2}_n - 
x(1-x) q^2}{m^2_n - x(1-x)q^2} \right]\ .
\eea

\begin{figure} [htb!]
\begin{center}
\epsfig{file=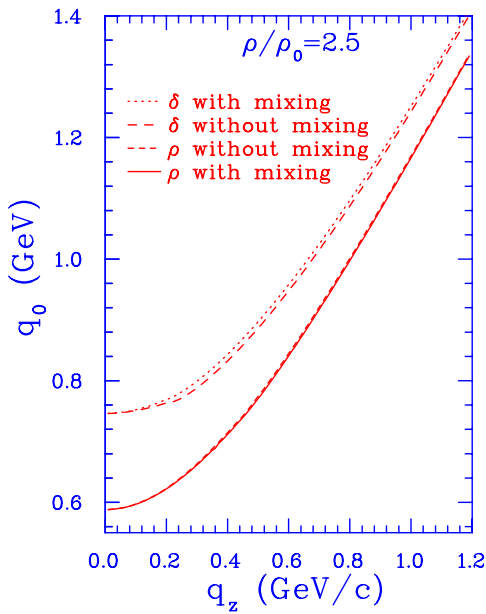,height=8.0cm,angle=0}
\end{center}
 \caption[Dispersion curve] 
  {\small The dispersion curve for $\rho$ and $\delta$ meson with and
  without mixing at ${\var\rho}$=2.5${\var\rho}_0$. \label{fig:dispersion}}
\end{figure}

It might be worthwhile to say here that the $\rho$ meson, being a vector,
the collective oscillations set by its propagation through matter would have 
longitudinal (L) and transverse (T) components 
depending upon whether its spin is aligned along or perpendicular to the
direction of propagation. Accordingly, with a special choice of z-axis along
the direction of the momenta $({\vec q})$, one can define the longitudinal and 
transverse polarization as $\Pi_L= - \Pi_{00} + \Pi_{33} $ and 
$\Pi_T=\Pi_{11}=\Pi_{22}$ respectively \cite{chin77}. 
To determine the collective modes, one defines the
dielectric function as \cite{chin77}:
\bea
\epsilon(q_0,|{\vec q}|) &=& \det (1 - {\cal D}^0 \Pi)
   = \epsilon_T^2 \times \epsilon_{mix}\ ,
\eea
where $\epsilon_{T}$ corresponds to
two identical transverse (T) modes and $\epsilon_{mix}$ correspond to the
longitudinal mode with the mixing. The latter, of course, also characterizes
the mode relevant for the $\delta$ meson propagation.
\bea
\epsilon_T &=& 1 - d_0 \Pi_T, \hspace{0.6cm}
d_0 = \frac{1}{q^2 - m_{\delta}^2 + i\epsilon}\label{trns} \nonumber\\
\epsilon_{mix} &=& (1 - d_0 \Pi_L)(1 - \Delta_0 \Pi_s) -
\frac{q^2}{|{\vec q}|^2} \Delta_0 d_0 (\Pi_0)^2 
\label{long}
\eea
The zeros of the dielectric functions 
characterize  the dispersion relation for the meson propagation.
Fig.~\ref{fig:dispersion} shows the relevant dispersion curves with
and without mixing at density ${\var\rho}$=2.5${\var\rho}_0$. 
As only the L mode mixes with the scalar mode, we do not consider the
T mode. The latter in fact is the same as presented in Ref. \cite{abhee97} for
$\rho$ meson.  The effect of mixing on the pole masses, as evident from
Fig. \ref{fig:dispersion}, are found to be very small. However, the mixing 
could be large when the mesons involved go off-shell. It should be
noted that the modes with mixing  move away from each other
compared to what one obtains without mixing. This
can be understood in terms of ``level-level'' repulsion
driven by the off-diagonal terms of the dressed propagator
\cite{saito98}.

\begin{figure}[htb!]
\begin{center}
\epsfig{file=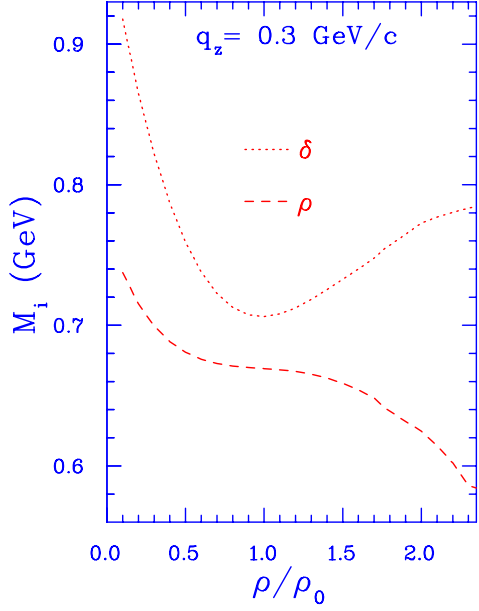,height=8.0cm,angle=0}
\end{center}
 \caption
  {\small The invariant mass as a function of the relative nuclear
density.  \label{fig:invmass}}
\end{figure}

Fig. \ref{fig:invmass} shows the dependence of the invariant masses
($M_i = {\sqrt{q_{0(i)}^2-|{\vec q_i}|^2}}$)
( i = $\rho, \delta$ ) on nuclear densities where $q_0$'s are
determined from the zeros of the dielectric function, Eq. (\ref{long}). 
It is evident that the difference of the invariant masses first
decreases with density reaching a minima and then again starts
increasing. This behavior arises from the non-monotonic density
dependence of the polarization functions.

\begin{figure}[hbt!]
\begin{center}
\epsfig{file=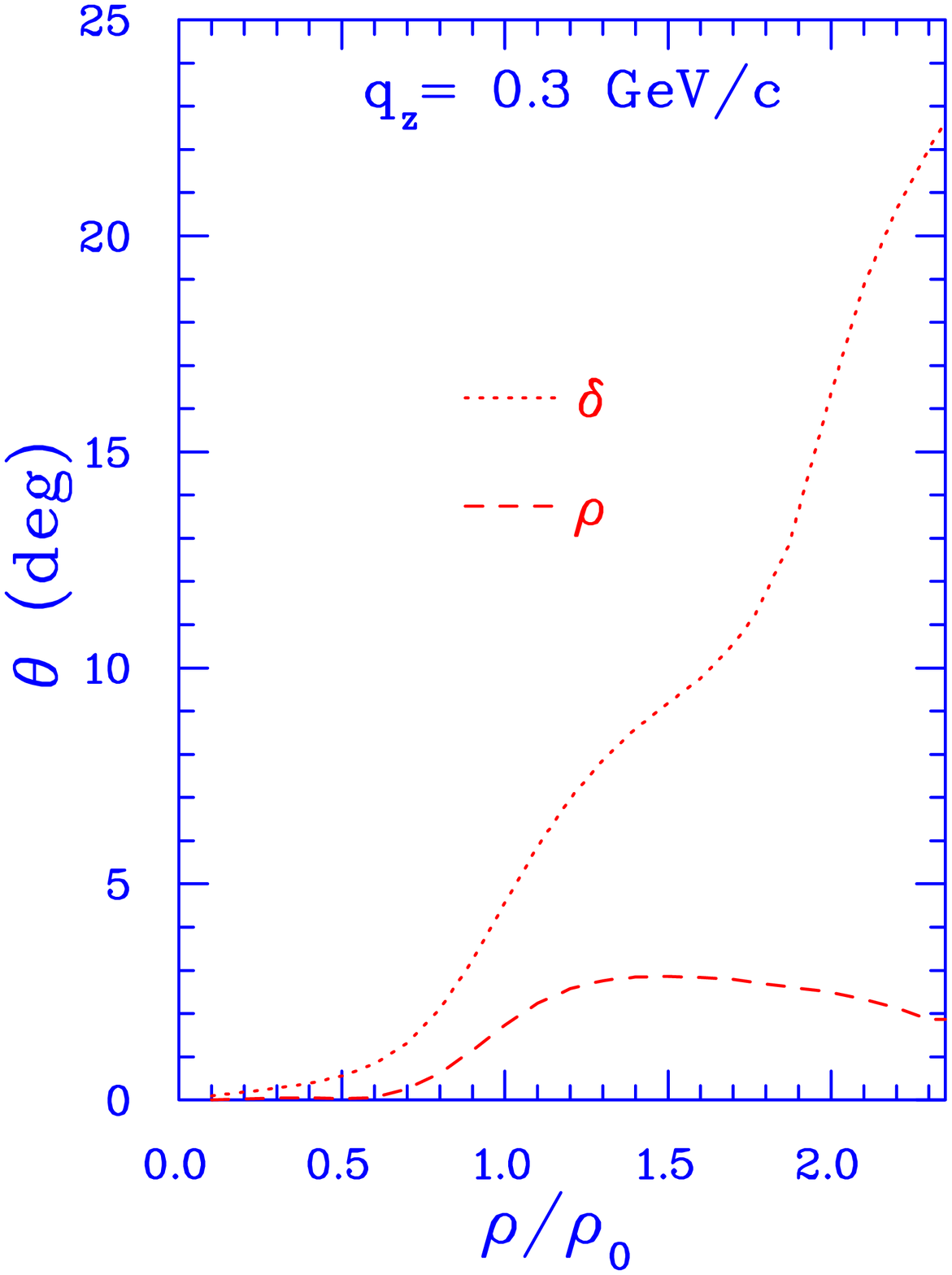,height=8.0cm,angle=0} \epsfig{file=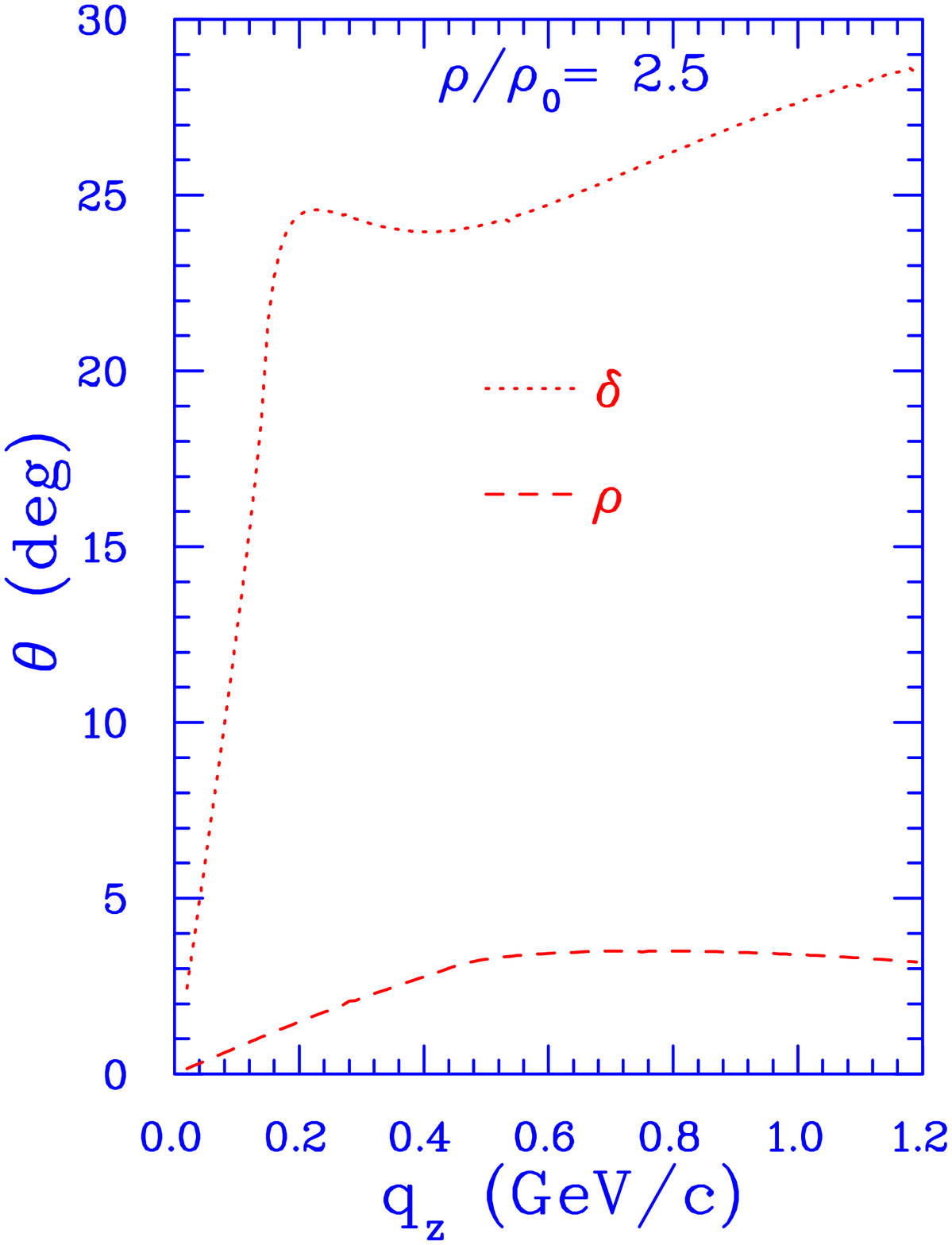,height=8.0cm,angle=0}

\end{center}
 \caption[Mixing angle] 
  {\small The mixing angle as a function of the relative nuclear density
(left panel) and of momentum (right panel).  \label{fig:angle}}
\end{figure}

To calculate the mixing angle, one diagonalizes the mass matrix\cite{abhee97} 
with the mixing and obtains
\bea
\theta_{mix} = \frac{1}{2} \arctan (\frac{2 \Pi^{\rho\delta}_{mix}}{
m_\delta ^2 - m_\rho^2 - \Pi_L^\rho + \Pi^\delta })
\label{mixangle}
\eea
In Eq.~(\ref{mixangle}) $\Pi^{\rho\delta}_{mix}= M_i/|{\vec q}|\Pi_0$ which
increases with density. $\Pi_0$ is the zero component of 
Eq.~(\ref{mixamp}).  Eq.~(\ref{mixangle}) clearly shows that the 
mixing angle depends not only on the mixing amplitude ($\Pi_0$) but
also on the ``energy denominator''. The latter, as seen in 
Fig. \ref{fig:invmass}, first decreases as a function of density then
again shows an increase characterizing the density dependence of the
mixing angle as presented in Fig. \ref{fig:angle}.
The mixing angle in Fig.~\ref{fig:angle} corresponds to $|{\vec q}|$=0.3 
GeV/c.
The momentum dependence for a density
2.5 times higher than the normal nuclear matter density is shown in
the right panel of the same figure. This shows that for momenta beyond 
$|{\vec q}|\approx 0.2 $ GeV/c the mixing is quite appreciable which
affects the dilepton yield substantially as shown later. It should
also be noted that the mixing angle vanishes at
$|{\vec q}|$ = 0 or at $\rho$ = 0, as it should.

\begin{figure} 
\begin{center}
\epsfig{file=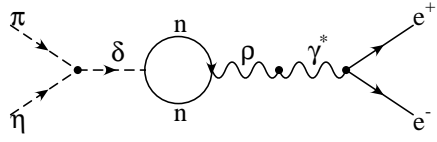,width=6.0cm,angle=0}
\end{center}
 \caption[Loop] 
  {\small The Feynman diagram for the process $\pi + \eta 
\rightarrow e^+ + e^-$. \label{fig:loop}}
\end{figure}

The $\rho$-$\delta$ mixing opens a
new channel {\em viz} $\pi + \eta \rightarrow e^+ + e^-$ in dense
nuclear matter through n-n excitations. The Feynman diagram of the
process is  depicted in Fig.~\ref{fig:loop}. 

The cross-section for this
process might be expressed in terms of the mixing amplitude ($\Pi_0$)

\bea
\sigma_{\pi \eta \rightarrow e^+ e^-} = \frac{4 \pi \alpha^2}{3 q_z^2 M}
\frac {g_{\delta\pi\eta}^2}{g_\rho^2}
\frac{m_\rho^4}{(M^2 - m_\rho^2)^2 + m_\rho^2 \Gamma^2_\rho(M)}
\frac{1}{(M^2 - m_\delta^2)^2 + m_\delta^2 \Gamma^2_\delta(M)}
\frac{1}{{\sqrt{M^2 - 4 m_\pi^2}}} {\mid\Pi_0\mid}^2 \ .
\eea

For the decay widths we consider the
invariant mass dependence as presented below:
\bea
\Gamma_\rho(M)=
\frac{g_{\rho\pi\pi}^2}{6\pi}
\frac{(\frac{M^2}{4}-m_\pi^2)^{\frac{3}{2}} \ ,
}{M^2} 
\eea

and
\bea
\Gamma_\delta(M)=
\frac{g_{\delta\pi\eta}^2}{16\pi}
\frac{{\sqrt{(M^2-(m_\pi+m_\eta)^2)(M^2-(m_\pi-m_\eta)^2)
}} }{M^3} \ .
\eea

To describe the $\pi\delta\eta$ vertex we use  

\bea
{\cal L}_{\delta\pi\eta}=f_{\delta\pi\eta}\frac{m_\delta^2-m_\eta^2}{m_\pi}
\phi_\eta{\vec \phi}_\pi\cdot{\vec\phi}_\delta\ ,
\eea
where for later convenience we define $g_{\pi\delta\eta}=f_{\delta\pi\eta}
(m_\delta^2-m_\eta^2)/m_\pi $. 
\begin{figure}
\begin{center}
\epsfig{file=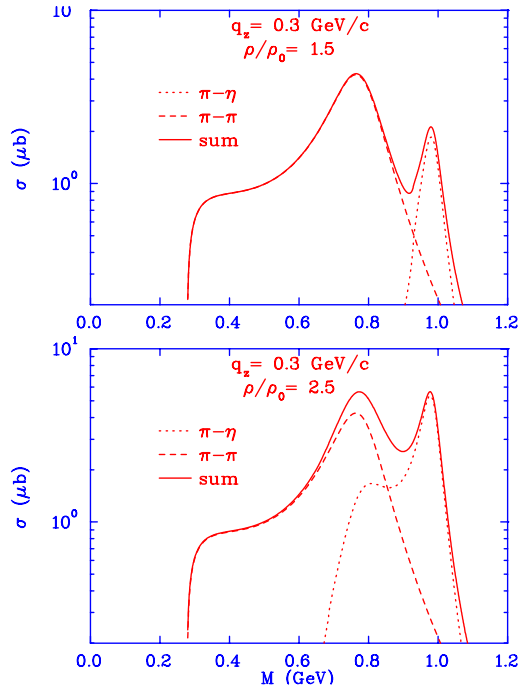,height=9.0cm,angle=0}
\end{center}
 \caption[Dilepton spectrum] 
  {\small Dilepton spectrum induced by $\pi + \pi
\rightarrow e^+ + e^-$ and $\pi + \eta
\rightarrow e^+ + e^-$ considering matter induced $\rho-\delta$ mixing \label{fig:dilepton}}
\end{figure}
Of course, there is an uncertainty involved with the coupling parameter
$f_{\delta\pi\eta}$ as discussed in Refs.\cite{kirch96,pdb}. This arises 
from the fact that 
$\delta$ (or $a_0$) lies close to the opening of the ${K{\bar K}}$ channel 
leading to a cusp-like behavior in the resonant amplitude, therefore a naive
Breit-Wigner form for the decay width is inadequate. 
Furthermore, as mentioned before, there is also uncertainty involved with
the $\delta$NN coupling which renders the precise extraction of $\delta$-
$\pi$-$\eta$ coupling even more difficult \cite{kirch96}.
We take a value for $f_{\delta\eta\pi}$=0.44 from Ref.\cite{kirch96} which
gives $\Gamma_{\delta\rightarrow \pi \eta}$($m_\delta$)=$59$~MeV,
while the experimental vacuum width of $\delta$ is 
between $50-100$~MeV \cite{pdb}.

One can notice in Fig. \ref{fig:dilepton} that the  process, 
$\pi + \eta \rightarrow e^+ + e^-$, at densities higher than ${\bf \rho_0}$,
not only enhances the overall production of lepton pairs but also
induces an additional peak near the $\phi$ mass region. The contribution
at the $\delta$ mass is comparable to that of  
$\pi + \pi \rightarrow e^+ + e^-$ near the $\rho$ peak, 
for densities higher than ${\bf \rho_0}$. 
Fig.~\ref{fig:dilepton} also shows that as the density goes even higher the
dilepton yield arising out of the mixing also increases further. The 
cross-section increases with increasing momenta of the mesons in
keeping with the mixing angle as shown in Fig.~\ref{fig:angle}.

We have highlighted the possibility of 
$\rho$-$\delta$ mixing in dense nuclear matter. 
We observe the appearance of an additional peak at a dilepton invariant
mass that corresponds to that of the $\delta$. With sufficient
experimental resolution, this effect could be observable. Probably not as
an individual peak, because of the $\delta$'s vacuum width which is
already not small, but more realistically as a shoulder in the $\phi$
spectrum. This feature is then exclusively density-dependent. 
Our aim here was to establish the existence of the signal.
Our calculation can, and will be
improved upon: further studies 
are in progress to assess finite temperature effects and to
self-consistently incorporate the necessary many-body 
machinery. For example, the characteristics of the $\rho$ can be
modified in the nuclear medium \cite{rw} and the in-medium behaviour of
the $\delta$ needs to be addressed. We have verified that the 
inclusion of hadronic form
factors does not change the conclusions we reach in this work. Detailed results will be presented elsewhere.

\acknowledgements                                

This work was supported in part by the Natural Sciences and Engineering
Research Council of Canada and in part by the Fonds FCAR of the 
Qu\'ebec Government.

%\end{multicols}
\end{document}